\documentclass[twocolumn]
{aastex631}
\pdfoutput=1

\begin{document}

\title{How Many Bursts Does it Take to Form a Core at the Center of a Galaxy?}

\author[0009-0009-0239-8706]{Olivia Mostow}
\affil{Department of Astronomy, University of Virginia, 530 McCormick Road, Charlottesville, VA 22903, USA}

\author[0000-0002-5653-0786]{Paul Torrey}
\affil{Department of Astronomy, University of Virginia, 530 McCormick Road, Charlottesville, VA 22903, USA}

\author[0000-0002-2628-0237]{Jonah C. Rose}
\affiliation{Department of Astronomy, University of Florida, Gainesville, FL 32611, USA}
\affiliation{Department of Physics, Princeton University, Princeton, NJ 08544, USA}
\affiliation{Center for Computational Astrophysics, Flatiron Institute, 162 5th Avenue, New York, NY 10010, USA}

\author[0000-0002-8111-9884]{Alex M. Garcia}
\affiliation{Department of Astronomy, University of Virginia, 530 McCormick Road, Charlottesville, VA 22903, USA}

\author[0009-0002-1233-2013]{Niusha Ahvazi}
\affiliation{Department of Astronomy, University of Virginia, 530 McCormick Road, Charlottesville, VA 22903, USA}

\author[0000-0002-8495-8659]{Mariangela Lisanti}
\affiliation{Center for Computational Astrophysics, Flatiron Institute, 162 5th Avenue, New York, NY 10010, USA}
\affiliation{Department of Physics, Princeton University, Princeton, NJ 08544, USA}

\author[0000-0002-3204-1742]{Nitya Kallivayalil}
\affiliation{Department of Astronomy, University of Virginia, 530 McCormick Road, Charlottesville, VA 22903, USA}

\begin{abstract}
We present a novel method for systematically assessing the impact of central potential fluctuations associated with bursty outflows on the structure of dark matter halos for classical and ultra-faint dwarf galaxies. 
Specifically, we use dark-matter-only simulations augmented with a manually-added massive particle that modifies the central potential and approximately accounts for a centrally-concentrated baryonic component. 
This approach enables precise control over the magnitude, frequency, and timing of
rapid outflow events.
We demonstrate that this method can reproduce the established result of core formation for systems that undergo multiple episodes of bursty outflows. 
In contrast, we also find that equivalent models that involve only a single (or small number of) burst episodes do not form cores with the same efficacy.
This is important because many UFDs in the Local Universe are observed to have tightly constrained star formation histories that are best described by a single, early burst of star formation. 
Using a suite of cosmological, zoom-in simulations, we identify the regimes in which single bursts can and cannot form a cored density profile.
Our results suggest that it may be difficult to form cores in UFD-mass systems with a single, early burst regardless of its magnitude.
\end{abstract}


\section{Introduction} \label{sec:intro}

The current paradigm of cold, collisionless dark matter plus dark energy ($\Lambda$CDM) has had a number of successes reproducing observations on the largest scales.
For example, the distribution of galaxy clusters seen from large sky surveys are consistent with the predictions of cosmological N-body simulations which evolve the primordial fluctuations measured from the Cosmic Microwave Background to present-day using a $\Lambda$CDM framework~\citep[e.g.,][]{Springel2005, Reid2010}.
Yet, on smaller scales (i.e., the scales of individual galaxies), key tensions persist that bring into question whether a cold and collisionless dark matter (DM) species is indeed the best descriptor~\citep{Bullock2017, Sales2022}.

The small-scale tensions in Cold Dark Matter~(CDM) generally arise from inconsistencies in the predicted structure, abundance, or distribution of galaxies as compared to observations---especially for low-mass systems~\citep{Moore1994,Sales2022}. In the absence of baryons, CDM galactic halos are expected to follow Navarro-Frenk-White~(NFW)  profiles~\citep{NFW}, which feature a continuously rising central DM density (i.e., a DM cusp) with an inner log-slope $\frac{dlog(\rho)}{dlog(\rm r)}$ that approaches $-1$ at small radii.
By contrast, it is observationally inferred that some nearby low-mass galaxies have a flattened inner DM density (i.e., a DM core)~\citep[e.g.,][]{Walker2011, Oh2015, Almeida2024, Vitral2024}.  This discrepancy between predictions and observations is referred to as the ``core-cusp problem.''  It is related to the ``diversity problem''~\citep{Oman2015}, which refers to the fact that the circular velocity profiles of simulated CDM galaxies exhibit less variation compared to the observed profiles of dwarf galaxies.  This inconsistency may be an indication that the CDM model that underpins our current cosmological paradigm needs revisiting.
Indeed, alternative DM models, such as those which include self interactions~\citep[e.g.,][]{Spergel2000, Tulin2018}, can potentially alleviate these discrepancies by introducing new mechanisms that alter a halo's inner structure.

However, it is worth noting that the core-cusp problem was initially identified and studied through N-body simulations, which lack a direct treatment of the baryonic component of galaxies.  Modern cosmological simulations include not only gravity acting upon DM, but also hydrodynamics coupled to comprehensive models of galaxy formation. These simulations incorporate the physics of gas cooling, stellar feedback, AGN feedback, star formation, black holes, and the ISM~\citep[see, e.g.,][for reviews]{Somerville2015, Vogelsberger2020}.
The inclusion of baryons is not just an improvement allowing for more direct modeling of the emergent galaxy population, but is also critical for the evaluation of the CDM tensions as the baryons can impact the DM particles by modifying the overall halo potential.


In fact, it has been demonstrated that the interaction of DM and baryons through gravity alone may be sufficient to alleviate the core-cusp problem~\citep{Navarro1996, Read2005, Governato2012}.
Episodic mass ejection from the galactic center can inject heat into the central DM by rapidly fluctuating the central potential~\citep{Governato2010, Pontzen_Governato_2012, Pontzen_Governato_2014}.
Physically, these episodic mass ejections can be thought of as strong, or ``bursty'', stellar feedback events.
Fully cosmological simulations employing the more bursty feedback models have shown core formation beginning in the classical dwarf ($M_* = 10^{5}$--$10^{7} ~M_\odot$) regime and peaking in strength for bright dwarf ($M_* = 10^{7}$--$10^{9} ~M_\odot$) galaxies \citep[][]{DiCintio2014, Chan_2015, Tollet_2016, Bullock2017, Lazar_2020, Azartash-Namin2024}.
Thus, if star formation occurs in stochastic bursts~\citep[e.g.,][]{Governato2010, Hayward2017, FaucherGiguere2018}, as opposed to the stellar mass growing smoothly over time, the resulting gaseous outflows naturally perturb the orbits of the DM particles in the inner region and sufficiently flatten the density profile~\citep[e.g.,][]{Onorbe2015, Jahn2023}.
There are, however, limits where stellar feedback is incapable of DM core formation.
\cite{Fitts_2017} found $M_* \approx 2 \times 10^{6} ~M_\odot$ to be the threshold mass for bursty feedback to significantly modify the DM density profile. 
This limit is primarily, if not entirely, an energetic one: galaxies that are too low in mass simply do not have enough stellar feedback energy present to convert the core into a cusp.


Regardless of the details, it is clear that the small-scale tensions may arise either owing to our lack of knowledge of galaxy formation physics or a fundamental flaw in our currently favored DM paradigm.  
Studying both (i) the regimes when, where, and how bursty feedback is able to operate and, separately, (ii) the regimes where bursty feedback is able to convert DM cusps into cores is therefore critical to our understanding of the extent to which our DM prescription needs to be modified.
In this work, we take a new approach of using fully cosmological, dark-matter-only~(DMO) simulations coupled to analytically-modulated central potential contributions.  
We are able to capture the fully cosmological development of the DM halo, while also considering how \textit{systematically varied} time-dependent central potentials impact the DM halo structure.
As we demonstrate within this paper, we can systematically vary the total number of bursts, as well as the total mass ejected in each burst.  
Our approach allows us to fill in an important void that sits between previous idealized studies (e.g., those of \citeauthor{Pontzen_Governato_2012} \citeyear{Pontzen_Governato_2012} and \citeauthor{Ogiya_Mori_2014} \citeyear{Ogiya_Mori_2014}) and the fully cosmological studies where the bursty nature of feedback naturally arises in a way that cannot be directly modulated \citep[e.g.,][]{Hopkins2014, Onorbe2015, Chan_2015, Lazar_2020}.


In addition to introducing this flexible framework, we also aim to address the question: ``Could galaxies convert a DM cusp into a core via a single episode of star formation?''
It has been suggested that some UFD satellite galaxies have implied cored DM density profiles~\citep{Almeida2024}.
However, the stellar populations within these systems are fairly tightly constrained to be consistent with approximately single-age stellar populations that are also very old (i.e., having formed $> 80$\% of their stellar mass before the midpoint of reionization ($z = 7.7 \pm 0.7$; \citeauthor{Sacchi2021} \citeyear{Sacchi2021}).
In other words, if these systems formed most of their stellar mass in a single burst long ago, could that be sufficient to convert a DM cusp into a core, or are multiple, episodic bursts required?

In this paper, we test the ability of bursty feedback to convert cusps into cores using modified cosmological simulations where we can manually prescribe the number and magnitude of the bursts to address this question.
The structure of this paper is as follows.
In \S\ref{sec:Methods}, we outline our methods, including descriptions of: our simulations (\S\ref{sec:Simulations}), the employed galaxy growth/burst/outflow models (\S\ref{sec:GrowthModels},  \S\ref{sec:GalaxyModels}), and the method of characterizing inner DM density profiles (\S\ref{sec:DensityProfiles}).
In \S\ref{sec:Results}, we present our results, split into the classical dwarf (\S\ref{subsec:ClassicalDwarfs}) and UFD (\S\ref{subsec:UltraFaintDwarfs}) regimes.
In \S\ref{sec:Discussion}, we discuss our results and in \S\ref{sec:Conclusions}, we present our conclusions.

\section{Methods} \label{sec:Methods}

\subsection{Simulations}
\label{sec:Simulations}
We study the mechanism of core formation via bursty feedback using modified cosmological simulations.  
The foundation of our cosmological simulations are standard DMO zoom-in simulations.  
We create zoom-in initial conditions using \textsc{music}~\citep{Hahn2011}, with a parent box of 36~Mpc. We adopt cosmological parameters $\Omega_0 = 0.301712$, $\Omega_{\Lambda} = 0.6983$, $\Omega_b = 0.0$, and $H_0 = 100~\rm {h~km~s^{-1}}$ where $h = 0.6909$ consistent with Planck Collaboration XI \citep{Planck2016}. 
Our simulations have a DM mass resolution of $3.44 \times 10^{3} M_{\odot}$ and a gravitational softening of 0.038~kpc.
We then evolve these initial conditions from redshift $z=127$ down to $z=0$ using the simulation code \textsc{arepo}~\citep{Springel2010, Weinberger2020}.
Prior to the addition of any changes to the central potential (as described below), the setup described here is a standard CDM cosmological zoom-in simulation.

Our goal in this paper is to probe the impact of a set of successive mass expulsion events on the resulting DM halo structure.
To achieve this, we follow the approach used in~\cite{Rose2023} whereby the gravitational impact of the central baryon component (i.e., galactic gas and stars) is  represented by a single massive simulation particle. 
Unlike the DM simulation particles, which move freely under the force of gravity through the simulation domain and have a constant mass with time, the tracer particle is pinned to the potential minimum of the halo and given a manually prescribed mass that evolves with time. 
As with all other simulation particles, the tracer particle is assigned a gravitational softening.  
While this is strictly to avoid two-body interactions for DM particles, a larger softening is employed for the massive tracer particle to emulate the effect of having a spatially distributed (i.e., not point-like) mass distribution.
We note that a different potential modification profile could have been selected, however, ~\cite{Rose2023} showed this was sufficient to produce realistic galaxy properties when compared with baryonic simulations. 

We note four points about this setup.
First, because we manually vary the tracer particle's mass, the total mass in the simulation is not conserved.  
This variation in the particle's mass can capture the impact of baryons condensing into the central region and subsequently being expelled. 
In other words, this particle is the mechanism that we use to impose specific time variability in the central potential.
We do not expect the fact that the global mass budget changes slightly with time to impact our findings.
While local fluctuations to the mass (i.e., bursty outflows) can change the distribution of matter in the central halo, these fluctuations amount to a change in the total mass globally that is of order $10^{-6}\%$. 
Second, even though we are using a very simplistic method for modifying the central potential, the approach does capture the impact of the central baryon component on the DM halo structure, as demonstrated in~\cite{Rose2023}.
Third, because we are manually prescribing the tracer particle's mass, we can systematically explore varied mass growth histories including an array of smooth and bursty growth histories.
And, fourth, our controlled and simple approach remains agnostic to choices within the sub-grid modeling of baryonic processes (feedback prescriptions, star formation model, etc.). Instead, it focuses directly on the kinds of central potential fluctuations necessary to produce and sustain a DM core.
Even so, the trends we uncover by comparing the strength of core formation across this wide array of growth histories can potentially inform sub-grid modeling within future cosmological simulations by considering the kinds of ouflows driven in different model scenarios.
We discuss this further in the context of our results in \S \ref{sec:Discussion}.

We first explore the classical dwarf regime in order to ensure the model can reproduce previous results for the central DM halo properties in the presence and absence of bursty feedback.
We first model the smoothly forming classical dwarf galaxy  (i.e., without bursty feedback) to verify cusp formation. 
We then model the same system with the stellar mass forming over multiple episodic bursts of star formation.
Finally, we compare the resulting DM density profiles to verify that the tracer particle emulating bursty outflows can turn the cusp to a core.
Once this is established, we apply this method to the specific case of UFDs that undergo just one outflow.

\begin{figure*}[ht!]
    \centering
    \includegraphics[width=1.0\linewidth]{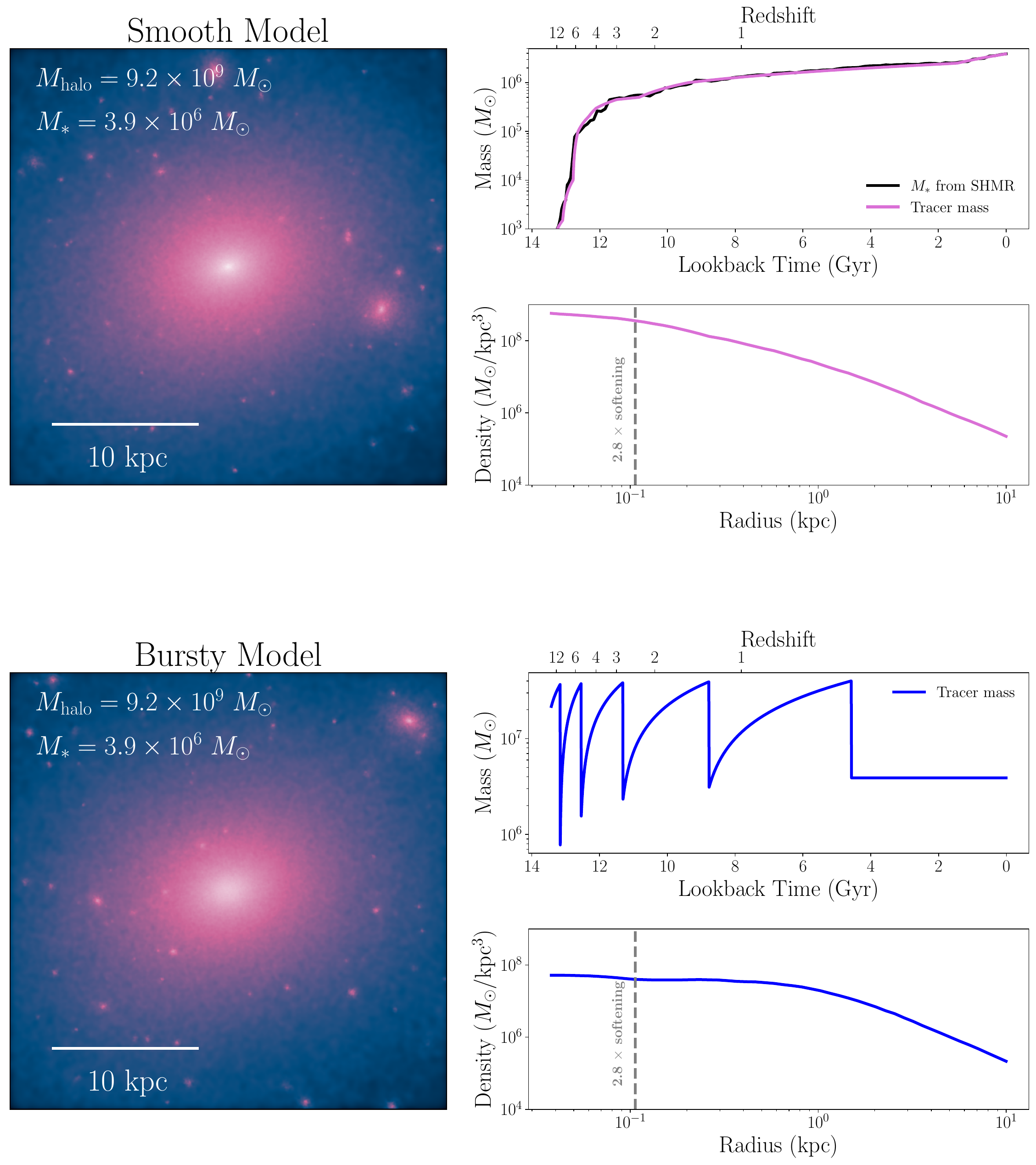}
    \caption{Comparison of DM density for the smooth~(top panels) and bursty~(bottom panels) models implemented using the modified-DMO approach for a halo with mass $9.2 
    \times 10^{9} M_{\odot}$. 
    \textit{Left:} Dark matter mass projection made from a $30 \times 30 \times 30$~kpc box surrounding the central halo. The halos look similar, save for their central regions. The smooth model produces a clear peak in density at the very center, while the bursty model reaches a constant density (forms a core) $2.05$~kpc from the center. 
    \textit{Right:} Tracer mass versus time as well as density profile for the smooth and bursty model. In both cases, the final stellar mass is $3.9 \times 10^{6} M_{\odot}$. The tracer mass in the upper panel represents a galaxy with a smooth history, which follows the SHMR shown in black. The tracer mass evolution in the lower panel represents a galaxy that undergoes five bursts of star formation, expelling $3.6 \times 10^{7} M_{\odot}$ of gas in each burst. The $z=0$ DM density profiles produced from each of these models are also shown. The grey dashed line marks $2.8\times$ the DM particle softening, the point below which numerical effects begin to be important. The smooth model has an inner log-slope $\alpha = -1.40$ when averaged over 1--2\% of the virial radius, consistent with an NFW profile. By contrast, when there are episodic bursts, we find a log-slope $\alpha = -0.63$.}
    \label{haloimg}
\end{figure*}

\subsection{Galaxy Growth/Outflow Models}
\label{sec:GrowthModels}
The unique feature of our simulations is that we impose specific mass evolution histories on the tracer particle.  Specifically, its mass is updated at each timestep during the simulation following some pre-prescribed pattern. 
We employ three growth patterns: a smooth model, an episodically bursty model, and a single-burst model---each of which are described here.

The first class of model is a smooth star formation history, henceforth referred to as smooth models, which describes a galaxy whose central mass smoothly grows with time (i.e., does not experience any significant mass blowouts).
The tracer mass growth over time for this model is shown in the top-right panel of Figure~\ref{haloimg}.
The primary purpose of these models is to provide a standard of comparison against which we can understand how the bursts have (or have not) impacted the density profile. 

The second class of model is the episodically bursty model, where the galaxy is assumed to accrete gas mass smoothly for some period of time, followed by a rapid/instantaneous drop in the tracer particle mass, mimicking a feedback-driven blowout event.
The tracer mass growth over time for one such model is shown in the lower-right panel of Figure \ref{haloimg}.
We keep the final stellar mass fixed at the same value as in the smooth model (described in further detail for each of the two galaxy models in \S \ref{sec:GalaxyModels}).
We prescribe a mass evolution for the tracer particle that is meant to mimic the fluctuating mass of a galaxy with bursty outflows and captures the impact that bursty feedback has on the orbits of the inner DM particles.
The fluctuation of the central mass changes the gravitational potential, and if this happens sufficiently quickly (not adiabatically), it boosts the DM particles to a higher orbit. 
The net effect is that this process irreversibly transfers energy to the DM particles. 
We model this process by allowing the tracer particle mass to increase at a constant rate in scale factor, then instantaneously decreasing it. 
The height of this drop physically corresponds to the gas mass expelled from the central region of the galaxy due to supernova energy. 
The mass that remains after this sharp decrease corresponds to the stellar mass formed in the burst (which is simply the final stellar mass divided by the number of bursts).
After the last burst, the tracer particle's mass remains constant.
This is not necessarily physical in the sense that the galaxy's mass can continue to increase, but is motivated by the fact that star formation is less bursty at lower redshifts~\citep{FaucherGiguere2018}. 
We vary both the number of bursts and the amount of mass expelled in \S \ref{subsec:ClassicalDwarfs} to understand the impact of both of these parameters. 
Rather than making physical assumptions, the range of values we tested for the amount of mass expelled was determined by starting with extremely large values and decreasing it until core formation ceased.


We assign the burst times by assuming that the time separating bursts is proportional to the dynamical time of the halo~\citep{Ogiya_Mori_2014}.
Episodically bursty models with more (fewer) bursts simply assume that the burst timing is a smaller (larger) multiple of the Hubble time.
In the model shown in the lower-right panel of Figure \ref{haloimg}, we assume that the stellar mass forms over five bursts (at $z=9.9,~5.3,~ 2.8,~1.3~{\rm and}~0.4$) and each burst causes $3.6 \times 10^{7} M_{\odot}$ to be expelled from the inner region of the galaxy.\footnote{One can easily see that the minimum mass after the burst is gradually increasing with time. In fact, the upper mass immediately before the burst is also increasing with time, but this is not so easily seen with the log-scale of the plot. This makes it seem like the amount of mass being expelled varies with time even though it is constant.}
The number of bursts that takes place in hydrodynamic simulations which model these processes explicitly can vary based on many factors, but is generally of order 10 bursts.
As one example, galaxies in a sample taken from the FIRE simulations underwent $\sim 1$--$2$ bursts every 200 Myr at $z=2$~\citep{Sparre2017}.
Approximating this as the rate across a galaxy's entire formation history, and assuming that star formation ceases to be bursty at $z \sim 1.3$~\citep{FaucherGiguere2018}, this yields a typical value of between 20 and 50 bursts.  

The third and final growth model we employ is the single-burst model. 
Inspired by stellar age distributions in UFDs~\citep[e.g.,][]{Sand2010, Okamoto2012, Weisz2014, Brown2014, Simon2019, Gallart2021, Sacchi2021}, these models increase their mass steadily with time (attributed to gas accretion) followed by a single outflow event.
The subsequent mass is then held constant until the present day. 
This model can be considered a limit of the bursty model that is constrained to a single burst.

\subsection{Galaxy Models: Classical and Ultra-faint Dwarf}
\label{sec:GalaxyModels}
We simulate both classical dwarf and UFD galaxies to study the mechanism of core formation via bursty feedback.
The classical dwarf has a $z=0$ halo mass of $9.2 \times 10^{9} M_{\odot}$.
To model a galaxy of this mass with a smooth growth history, we employ the redshift-dependent stellar-to-halo-mass relation~(SHMR) described in~\citet{Moster2013}. 
To achieve this, we first run a DMO simulation to calculate the halo mass as a function of time. 
We then use the SHMR to evaluate the corresponding stellar mass and prescribe the growth of the massive particle to match this growth history.
The tracer mass enters the simulation at $z=11$ with a mass of $10^{3} M_{\odot}$ and has a final mass of $3.9 \times 10^{6} M_{\odot}$, as shown in the top-right panel of Figure \ref{haloimg}.\footnote{The tracer particle is initialized at the same time of $z=11$ but at a higher mass ($2.3 \times 10^7 M_{\odot}$) for the bursty model, shown in the bottom-right panel of Figure \ref{haloimg}. This is because the first of the five bursts in this model is set to occur at $z=10$. In order for this to happen, the tracer mass must be quite large prior to the outflow at this time. 
}
We place the tracer particle in the simulation no earlier than $z=11$ because it will be pinned to the potential minimum of the most massive halo within a manually defined sub-volume. 
At very early times, the individual halos are closer together and more similar in mass, making it difficult to identify the most massive halo at $z=0$, which is where we want to place the tracer mass.
We do not expect this choice to significantly impact our results because the halo is a small fraction of its $z=0$ mass.
We use a gravitational softening of 0.36~kpc for the massive particle representing the baryon mass of the galaxy, meaning that the majority of the mass is concentrated within 1.1~kpc.

The UFD has a $z=0$ halo mass of $7.8 \times 10^{7} M_{\odot}$.
We adhere to observational constraints on the star formation histories of these systems to determine their stellar mass as a function of time.
We prescribe a growth history consistent with the cumulative mass fraction over time of Milky Way (MW) satellites that have a similar present-day halo mass. 
Specifically, we set the final stellar mass to $1.4 \times 10^{3} M_{\odot}$---consistent with the observed stellar mass of these systems, which ranges from $(0.54$--$7.46) \times 10^{3} M_{\odot}$~\citep{Sales2017} and within $2\sigma$ of both the predictions of semi-analytic models~\citep{Ahvazi2024} and forward modeling of MW satellites~\citep{Nadler2020}. 
In our smooth model, the particle mass grows such that 80\% of the stellar mass has formed by $z=7$ and 90\% by $z=5$. 
To model a single burst for this galaxy, rather than making physical assumptions about when the burst occurs, we manually vary its timing across a wide range of values ($z=1$ to $z=8$) to understand how this changes the impact of the outflow on the DM density. 
The amount of mass expelled is initially set at $1.4 \times 10^{4} M_{\odot}$ by making assumptions about the gas mass available to be expelled in these systems, the details of which are discussed in \S \ref{subsec:UltraFaintDwarfs}, but it is also a parameter that we vary.
We also employ a smaller gravitational softening of 0.06 kpc for the massive particle to account for the difference in size as compared with the classical dwarf system.


\subsection{Characterization of Density Profiles}
\label{sec:DensityProfiles}
We use two independent metrics to characterize the density profiles of the galaxies and assess whether they are cuspy or cored.
We calculate the inner log-slope, commonly referred to as $\alpha$ (where $\rho \propto r^{\alpha}$), over 1--2\% of the virial radius~\citep{DiCintio2014, Tollet_2016}.
This is a reliable metric for cores that are a few percent of the virial radius in size, but is not sensitive to the presence of smaller cores (given the radius range where $\alpha$ is determined). 
For this reason, we use a complementary approach of fitting to the core-Einasto model and finding the best-fit core radius $r_c$~\citep{Lazar_2020}. 
This addresses the issue of not detecting smaller cores, but there can still be instances in which the model fit is sufficiently poor that we do not get a reliable estimate of the core radius. 
Density profiles with slightly irregular shapes will not be well-characterized by a three-parameter model, and in these cases, the core radius estimate will also be an imperfect metric.
Using both of these measures decreases the bias associated with $\alpha$ or $r_{\rm {c}}$ alone.

\section{Results} \label{sec:Results}
In this section, we present results employing the prescriptive approach described in \S\ref{sec:Methods}.
We split our results into the classical dwarf regime (\S\ref{subsec:ClassicalDwarfs}), where DM cores are often observed~\citep[e.g.,][]{Kleyna2003, Walker2011, Amorisco2011, Amorisco2013}, and the UFD regime (\S\ref{subsec:UltraFaintDwarfs}) where the DM density may  be cored~\citep{Amorisco2017, Contenta2018, Simon2021} or cusped~\citep{Hayashi2020, Vitral2024},
and there are often large uncertainties associated with these measurements ~\citep{Hayashi2023}. 

\subsection{Classical Dwarfs}
\label{subsec:ClassicalDwarfs}

Figure~\ref{haloimg} provides a visual overview of the smooth (top row) and bursty (bottom row) growth models for the classical dwarf.
For each model, we show the mass growth of the tracer particle over time. 
For the smooth model, the tracer particle mass is shown in magenta against the stellar mass inferred from the SHMR relation in black~\citep{Moster2013}.
As described in \S \ref{sec:GrowthModels}, the episodically bursty model has a tracer mass that increases smoothly, then instantaneously decreases a specified number of times. In the model shown, there are five such outflow events. 
Also shown for each model are the corresponding present-day DM density profiles.
The vertical grey dashed line marks $2.8~\times$ the DM particle softening, the point below which numerical effects begin to be important.
Looking at the region to the right of this dashed line, we see that the central density profile is cusped for the smooth model and cored for the episodically bursty model---both consistent with expectations.
The smooth model has inner log-slope $\alpha = -1.40$ when averaged over $1$--$2\%$ of the virial radius, consistent with an NFW profile.
By contrast, the episodically bursty model forms a core $2.05$ kpc in size and has $\alpha = -0.63$.
For illustration, we also show the present-day DM mass projection, produced from a $30 \times 30 \times 30$ kpc box centered on each galaxy. 


\begin{figure*}
\includegraphics[width=1.0\linewidth]{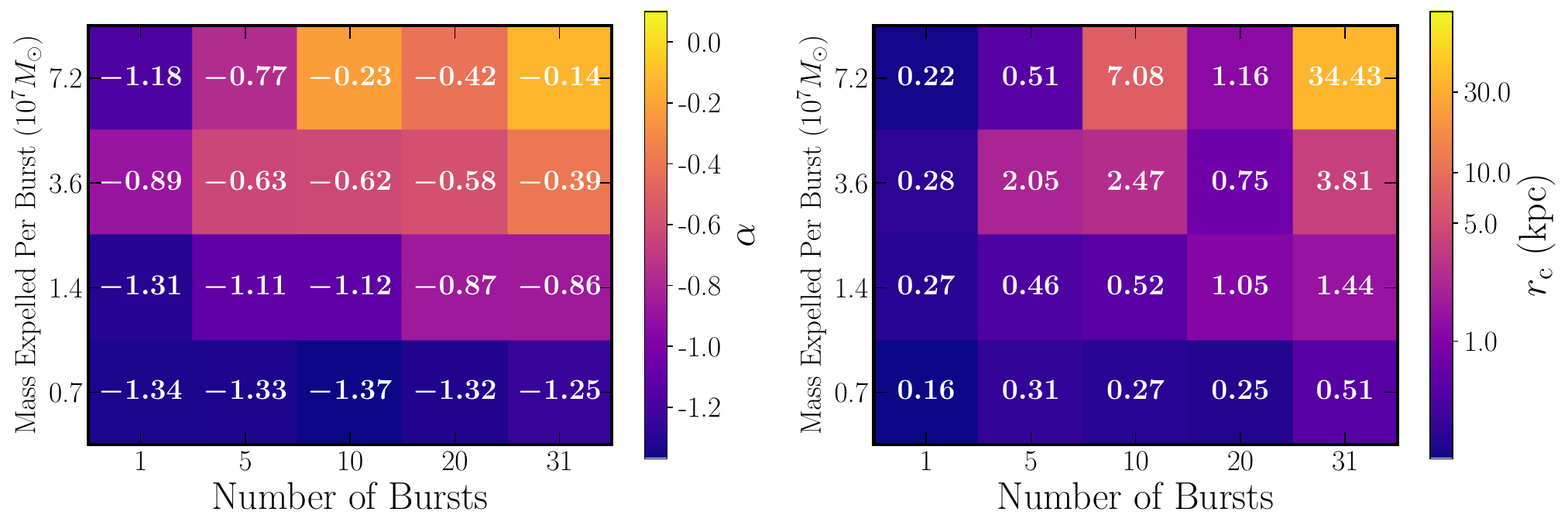}
    \caption{Summary of the inner log-slopes ($\alpha = \frac{\rm {dlog}\rho}{\rm dlogr}$) best-fit core radii $r_{\rm{c}}$ for a classical dwarf galaxy as a function of number of bursts and burst mass.}
    \label{mbgrid}
\end{figure*}

\subsubsection{Variable Total Mass Outflow Models}

As a first variation on the episodically bursty model, we consider how the number and amplitude of potential variations impact core formation.
Specifically, we manually change the strength and frequency of the central potential variations by altering (i) the total number of bursts ranging from 1 to 31, spaced proportionally to the dynamical time of the halo and (ii) the total mass expelled per burst ranging from $0.7\times10^7 \mathrm{M}_\odot$ to $7.2\times 10^7\mathrm{M}_\odot$.
Taken together, this changes the total integrated mass expelled for the tracer particle from $0.7\times10^7 \mathrm{M}_\odot $ (for a single burst with the smallest mass expulsion) to $2.2\times10^9~\mathrm{M}_\odot $ (for 31 of the most massive bursts).  
In other words, this tests the impact of changing the total mass ejected from the central region by several orders of magnitude---hence, we refer to these scenarios as variable total mass outflow models.

Our results show significant variations in core formation based on our outflow model. For each model, we have calculated both the inner log-slopes and best-fit core radii $r_{\rm{c}}$~\citep{Lazar_2020} of the resulting present-day DM density profiles. We show the results in terms of both metrics as a function of the amount of mass expelled per burst and number of bursts in Figure \ref{mbgrid}.
Several clear trends emerge: galaxies generally have more cored density profiles when they either experience a larger number of bursts (i.e., moving to the right in Figure~\ref{mbgrid}) or more massive bursts (i.e., moving up).
As the mass expelled per burst increases, the inner log-slope for the 10, 20, and 31 burst simulations increases from  $ - (1.37$--$1.25)$ to $-(0.42$--$0.14)$ monotonically. 
Similarly, the core radius for these simulations increases from 0.25--0.51~kpc to 1.2--34~kpc. 
We note that there is also some non-monotonicity, but that this can be mostly attributed to the density profile associated with these models having irregular inner shapes that are not fully described or well-characterized by the three-parameter core-Einasto model.
This trend continues to some extent for one and five bursts.
In these regimes, there is still a trend where more massive bursts lead to larger core radii, but the trend is somewhat less pronounced. While the 10, 20, and 31 burst simulations each increase the core radius by a factor of five or more, the single-burst model increases the core radius from 0.16--0.28~kpc---less than a factor of two. 
This result is unsurprising given that increasing the mass-per-burst directly increases the amplitude of the potential fluctuations.
Notably, the core radii we obtain for models with 10 bursts and varying mass expelled per burst of 0.27 to 7.08 kpc are comparable to those found in the FIRE-2~\citep{FIRE2} halos of similar mass in ~\citeauthor{Lazar_2020}~\citeyear{Lazar_2020} of 0.28 and 5.09 kpc which also contain bursty feedback (in which the stellar-to-halo mass ratio is increasing between the two rather than the size of the fluctuations directly as we model here). 
For any fixed value of the mass expelled per burst, the largest cores form for the greatest number of bursts. 
When the mass expelled is held constant at $3.6 \times 10^{6}M_{\odot}$, for example, the core radius increases from 0.28 to 3.81 kpc as the number of bursts increases from 1 to 31. 
Similarly, $\alpha$ increases from $-0.89$ to $-0.39$. 

\subsubsection{Constant Total Mass Outflow Models}

In contrast to the variable total mass outflow models considered in the previous subsection, here we hold the total mass ejected constant at $1.8 \times 10^{8} M_{\odot}$ while varying the number of bursts between 1--62 and mass-per-burst accordingly.
The total mass ejected is set to the value for the five-burst model shown in Figure \ref{haloimg} because that model has a cored present-day density profile. 
However, we note that there is no other reason to choose this particular value.

\begin{figure*}
\includegraphics[width=1.0\linewidth]{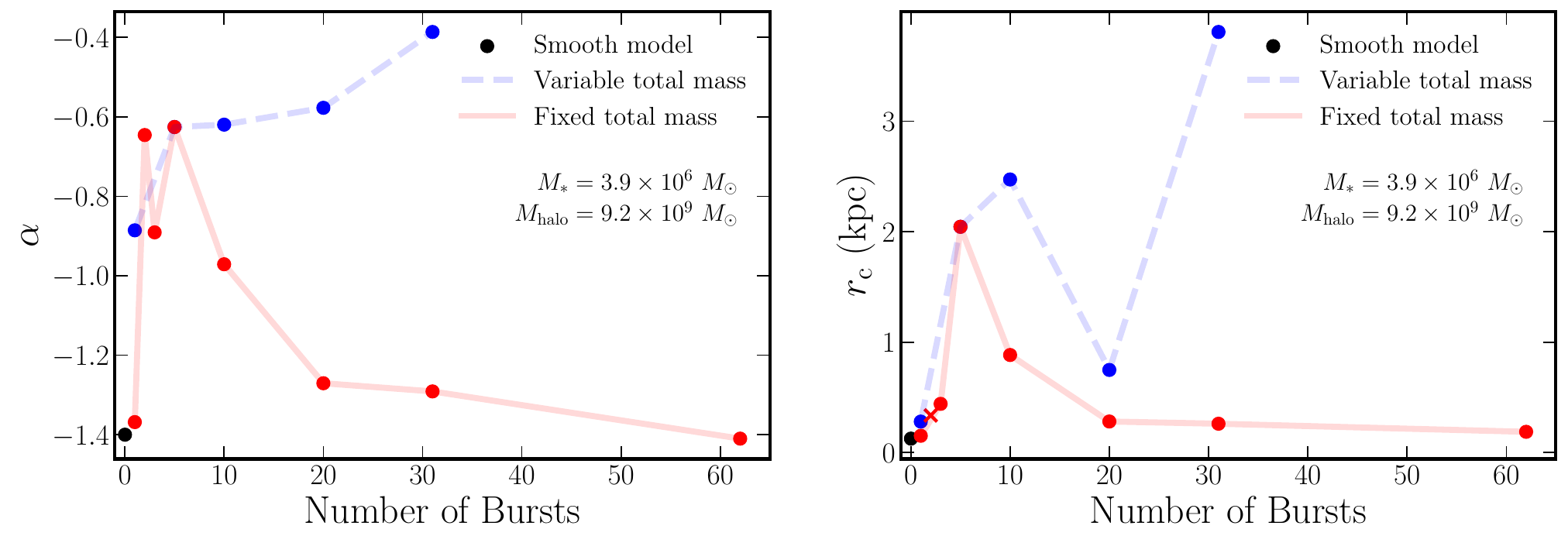}
    \caption{Comparison of the relationship between number of bursts and both the core size (right) and inner log-slope (left) for variable~(blue) versus fixed~(red) total energy transferred to the DM particles via bursty feedback for the classical dwarf galaxy. When the amount of mass expelled in each burst is constant, we find that the profile simply becomes more cored as we increase the number of bursts (increasing the total energy transferred). If instead the mass expelled in each burst is varied such that the total mass expelled is constant, we find a more complex relationship. As the number of bursts is decreased, initially this leads to an increase in the core size. However, if there are fewer than five bursts, this trend reverses. The point marked with an ``x" indicates that for this model, the core-Einasto fit is poor.}
    \label{varnum}
\end{figure*}
In Figure \ref{varnum}, we compare  the trends in core radii and inner log-slope for these constant total mass models with the variable total mass models of the previous section as a function of the number of bursts.
The blue line depicts the results when the total outflow mass varies with the number of outflows. Viewing the results in terms of the inner log-slope, we observe a monotonic upward trend from $-0.89$ to $-0.39$ as the number of bursts increases from 1 to 31.
Viewing the results in terms of the core radius, we find a similar trend where the size of the core increases from 0.28 to 3.81 kpc. 
The exception to this trend is the twenty-burst model, which has a smaller core radius than the ten-burst model. 
In this case, the density profile (not pictured) begins to flatten around 2 kpc, but the density increases again at smaller radii.
As a result, the shape of this profile is not well-fit by the three-parameter core-Einasto profile.
The trends observed with this line are simple to understand: more bursts of equal magnitude simply means more potential fluctuations capable of impacting the DM particle orbits.

\begin{figure}
	\includegraphics[width=\columnwidth]{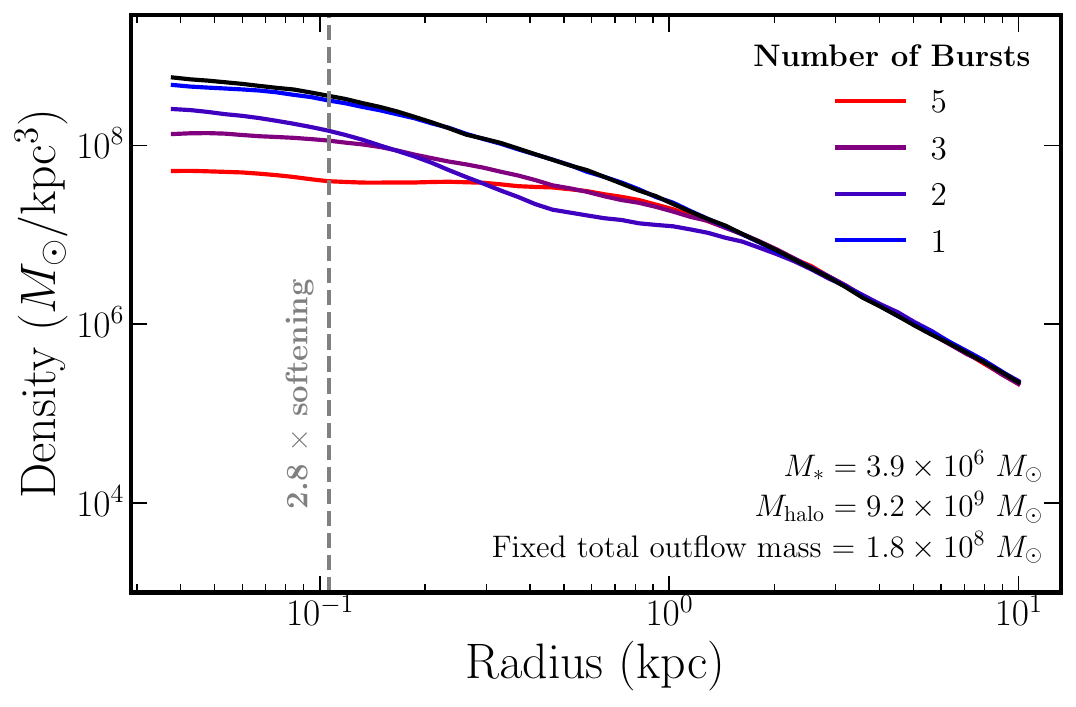}
    \caption{Present-day density profiles for a fixed amount of total mass expelled ($1.8 \times 10^{8}M_{\odot}$) over time and a varied number of bursts. The black line depicts the smooth model. As the number of bursts decreases (and therefore, the amount of mass expelled per burst increases) larger cores form.     However, cores stop forming altogether when the number of bursts becomes very small (i.e., $\sim 1$--2) despite the large amount of mass being expelled. The gray dashed line indicates the radius at which numerical softening effects begin to impact the results. Note that the results are shown for 1, 2, 3, and 5 bursts (not 4).}
    \label{fixedmass}
\end{figure}

The red line depicts the results when the total mass expelled is constant. 
The core radius begins at 0.15 kpc for one burst, increases to 2.05 kpc for five bursts, then monotonically decreases to 0.19 kpc.
The ``x'' indicates the one point for which a good fit is not obtained.
We observe a similar trend in $\alpha$ for these models: $\alpha=-1.37$ for one burst, which increases to and peaks at $-0.63$ for five bursts, then decreases to $-1.41$ for 62 bursts. 
The non-monotonicity in this trend from two to three bursts can be explained by looking at the density profiles shown in Figure \ref{fixedmass}. 
The two burst model has an inner log-slope and core radius of $\alpha = -0.65$ and $r_{\rm c} = 0.37$~kpc, but these values are more difficult to interpret due to the shape of the density profile.
From Figure \ref{fixedmass}, we can see that the density profile plateaus from 0.5 to 2 kpc, which overlaps with the range where the slope is calculated, but it begins to rise again for smaller radii. 
This means the calculated value of $\alpha$ does not completely characterize this density profile and explains the decrease in $\alpha$ from 2 to 3 bursts we see in Figure \ref {varnum}. 
In general, though, Figure \ref{varnum} shows that the dependence of core formation on number of bursts is more complex for the fixed total mass~(red) versus variable total mass~(blue) models.

Increasing the number of bursts makes a more cored profile initially, but this reverses after five bursts, which is perhaps unsurprising.
Indeed, in our fixed total outflow mass  models, the limit of $N_{\mathrm{bursts}} \rightarrow \infty $ becomes indistinguishable from a smooth model---which we have already demonstrated to form a cusp.  
What is more surprising is that core formation is mitigated in the limit where the number of bursts approaches only a single event (or very small number of events). 
To explore this further, we plot the the density profiles for the models with constant total outflow mass over 1, 2, 3, and 5 bursts in Figure~\ref{fixedmass}.  
The most cored profile is produced when the galaxy undergoes five bursts, shown in red, which has a core radius and inner log-slope 
of $2.05$~kpc and $-0.38$, respectively.
When there are three bursts, the density profile is steeper but still cored with $\alpha = -0.89$ and $r_{\rm c} = 0.44$~kpc.
The two and one-burst models, however, produce profiles that are more cusped.
As previously mentioned, the two-burst model has a shape that makes the slope shown in Figure \ref{varnum} less useful as a metric; we can see visually in Figure \ref{fixedmass} that it is not cored.
Finally, the one-burst model produces a profile nearly indistinguishable from the smooth model, with $\alpha = -0.83$ and $r_{\rm c} = 0.15$~kpc.

\subsection{Ultra-Faint Dwarfs}
\label{subsec:UltraFaintDwarfs}
As discussed in \S \ref{sec:intro}, some UFD galaxies have been observed (i) to have little variation in their stellar ages, and (ii) to have formed the vast majority of their stellar mass long ago~\citep[e.g.,][]{Brown2014, Simon2019, Gallart2021, Sand2010, Okamoto2012, Sacchi2021, Weisz2014}.
It has also been suggested that some of these systems have cored DM densities~\citep{Almeida2024}. 
Together, this begs the question: can a single, early burst turn a cusp to a core?
The approach we introduce in \S \ref{sec:GrowthModels} to modeling bursts of star formation allows us to shed light on this question by manually varying the timing and size of such a single outflow event to determine for what values of these parameters a core forms.

In all simulations described below, the galaxy has a final halo mass of $7.8 \times 10^{7} M_{\odot}$ and, to match observational mass measurements of the MW satellites that motivated this analysis, we prescribe a final stellar mass of $1.4 \times 10^{3} M_{\odot}$~\citep{Sacchi2021}. 
We assume that all of the ultra-faints have 95\% gas fractions prior to the burst, such that $M_{gas} = 1.4 \times 10^{4} M_{\odot}$. The pre-outflow mass ($M_{*} + M_{gas}$) is then equal to $1.54 \times 10^{4} M_\odot$. We assume all the gas is expelled in the outflow, leaving a post-outflow mass of $1.4 \times 10^{3} M_{\odot}$.\
Given that many of the UFDs in our Local Group have been observed to form $> 80\%$ of their stellar mass prior to reionization~\citep{Sacchi2021}, we begin by testing three single-burst models 
where the outflow from the galaxy occurs at $z=8,~7~{\rm and}~6$.
We compare the present-day density profiles of these models with that of a smooth model that also adheres to the aforementioned observational constraints on the stellar ages of these systems, as described in \S \ref{sec:GalaxyModels}.
The tracer mass in this smooth model is increased at each timestep such that 80\% of the stellar mass forms by $z=7$ and 90\% forms by $z=3$. 

\begin{figure}
	\includegraphics[width=\columnwidth]{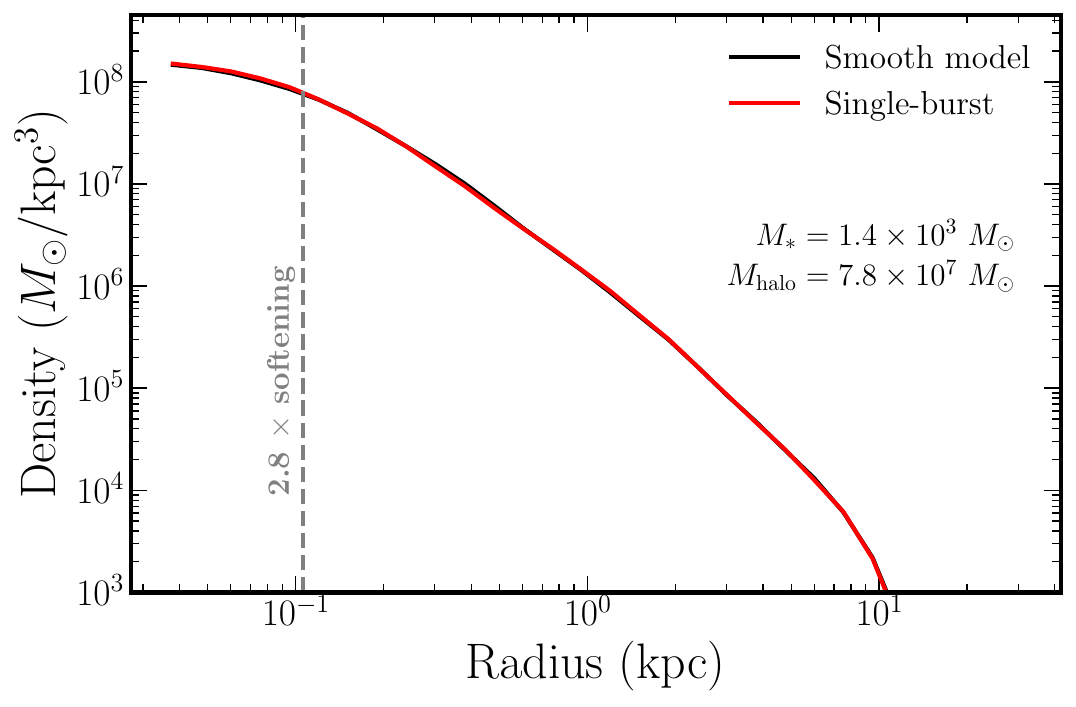}
    \caption{Comparison of the present-day DM density profile for an UFD galaxy if it has a smooth growth history (black) or forms in a single burst at $z=7$ (red) resulting in the ejection of $1.4 \times 10^{4} M_{\odot}$. The density profiles are both cusped with $r_{\rm c} = 0.07, 0.05 ~{\rm and }~ \alpha = -1.25,-1.35$ for the smooth-growth and single-burst models, respectively.}
    \label{singler7_m1}
\end{figure}

We find that none of the three single-burst models form a core, with $\alpha$ between $-1.35$ and $-1.15$ and $r_{\rm c}$ between 0.05 and 0.08~kpc.
As one example, Figure~\ref{singler7_m1} compares the density profiles of the smooth model and the single-burst model when the outflow occurs at $z=7$.
In the region to the right of the gray dashed line, which indicates the point at which numerical effects begin to impact the results, the two profiles are visibly cusped and nearly indistinguishable, as is the case for the models with an outflow at $z=6~\rm{or}~z=8$.



\begin{figure*}
    \includegraphics[width=1.0\linewidth]{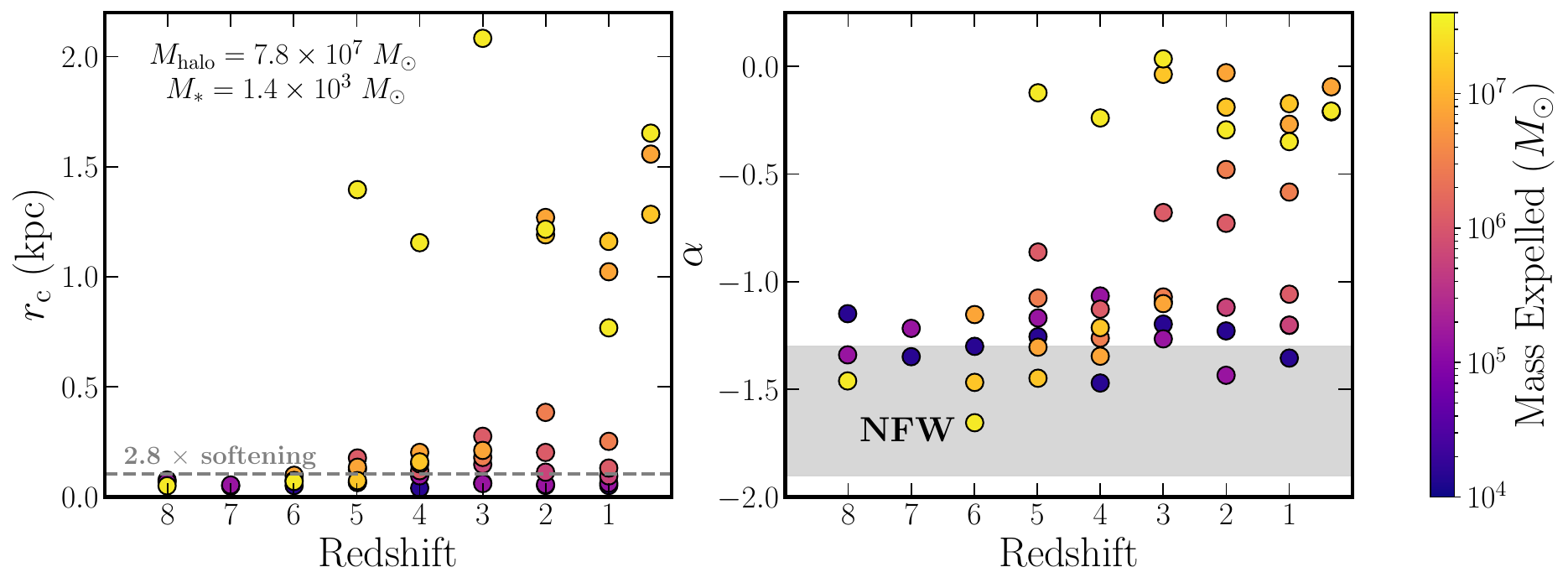}
    \caption{\textit{Left:} Best-fit values for the core radius $r_{\rm c}$ as a function of redshift at the time at which the single outflow occurs for the UFD. 
    The color depicts the size of the outflow in terms of the mass expelled.
    Notably, cores of appreciable size do not form prior to $z=5$, indicating that DM cores do not easily form when bursty activity is relegated to the earliest times, even when extreme mass outflow events are considered.
    \textit{Right:} Inner DM density profile slope, $\alpha$, as a function of redshift at which the single outflow occurs. 
    The shaded band represents the expected range of slopes for an NFW profile (accounting for concentration scatter).
    Just as with the left panel, the upper left region is unoccupied, showing consistency between these two metrics.}
    \label{m8single}
\end{figure*}

Finally, we run a suite of simulations to determine at what point, in terms of mass expelled and timing of the burst, cores begin to form.
The results, in terms of the inner log-slope and core radius, are summarized in Figure \ref{m8single} as a function of the time at which the burst occurs.
The horizontal dashed gray line in the left panel indicates the length scale affected by numerical softening.
A core radius below this should be disregarded, since the density profile is artificially flattened in this region.
The shaded gray band in the right panel indicates the expected range of values for an NFW~\citep{NFW} profile when accounting for concentration scatter found from N-body simulations~\citep{Maccio2007}, calculated using~\textsc{colossus}~\citep{colossus}.
We note that the mass-concentration relation from~\citeauthor{Maccio2007}~\citeyearpar{Maccio2007} was fit on more massive halos than the UFD in our simulations.
However, we include this shaded band not to make detailed quantitative comparisons with our results but rather to guide the eye. 
Whether we look at $r_{\rm{c}}$ or $\alpha$, it is apparent that outflows prior to $z=5$ impact the DM less than those at $z \ge 5$.
Prior to $z=5$, the profiles are all cusped, with negligibly small core radii relative to the softening length of the simulation particles and $\alpha < -1$. 
Outflows at $z\le5$, however, are able to impact the DM density for sufficient mass expelled.
The earliest outflow to form a core is that with $2.9 \times 10^{7} M_{\odot}$ expelled at $z=5$, with $\alpha = -0.12$ and $r_{\rm{c}} = 1.40$ kpc.
Less massive outflows, however, must occur later to have a dramatic impact on the present-day density.
As one example, an outflow of $1.2 \times 10^{6} M_{\odot}$ produces a cored density profile with $\alpha = -0.68$ and $r_{\rm{c}} = 0.28$ kpc if the burst occurs at $z=3$.
The same mass expelled at $z=6$ is less impactful, producing a density profile with $\alpha = -0.86$ and a negligibly small core radius.


\section{Discussion} \label{sec:Discussion}
In \S \ref{subsec:ClassicalDwarfs}, we find that our prescriptive approach to modeling bursty feedback can produce results that are consistent with literature expectations in the classical dwarf regime---by modeling 20 bursts we find cores of comparable size to those in 
the FIRE-2 simulations~\citep{FIRE2, Lazar_2020} 
and inner log-slopes comparable to those in the Storm suite (a subset of Marvel-ous Dwarfs; \citeauthor{Munshi2021}~\citeyear{Munshi2021}; \citeauthor{Azartash-Namin2024}~\citeyear{Azartash-Namin2024}).

In the latter study, the authors analyze the relationship between burstiness and core formation in dwarf galaxies across a wide range of masses and with two different sub-grid feedback prescriptions: the blastwave model~\citep{Stinson2006}, which models stellar feedback from individual supernovae, and the superbubble model~\citep{Keller2014}, which accounts for the fact that star formation is clustered through thermal conduction.
\citeauthor{Azartash-Namin2024}~\citeyear{Azartash-Namin2024} find that core formation takes place in bright dwarf ($M_* = 10^{7-10} M_{\odot}$) and some classical dwarf ($M_* = 10^{5-7} M_{\odot}$) galaxies 
and that core formation has a dependence on both $M_*/M_{\rm halo}$ and burstiness of star formation.
Notably, core formation was more strongly correlated with $M_*/M_{\rm {halo}}$ than burstiness in their superbubble model, which utilizes a more sophisticated model that accounts for supernova clustering and multiphase gas. 
While a direct comparison with this result is challenging since our parameters (mass expelled, number of bursts, etc.) do not neatly map onto the burstiness parameter that more naturally describes the hydrodynamic simulations within the Storm suite, our findings in \S \ref{subsec:ClassicalDwarfs} appear qualitatively consistent.
Provided there are more than three
outflows, we find that core formation is a stronger function of the amount of mass expelled than the number of outflow events.
Given that $M_*$ sets the amount of supernova
energy available to expel mass from the central region, the fact that~\citeauthor{Azartash-Namin2024}~\citeyear{Azartash-Namin2024} found a stronger dependence on this quantity than their burstiness parameter may be a manifestation of the same underlying principles as with our findings; the amount of mass expelled
determines the size of the gravitational potential fluctuations that impact the DM particle orbits.

Beyond yielding results consistent with the literature, the prescriptive and highly controlled method we employ here allows for a more thorough analysis of what is necessary and sufficient for bursty feedback to form cores in dwarf galaxies than is typically possible from hydrodynamic simulations.
The result of varying the total mass ejected from the galaxy (and therefore the total energy transferred to the DM particles) shown in Figure \ref{mbgrid} suggests perhaps how cusped or cored the present-day DM density is depends simply on the total outflow mass over all bursts.
However, we find that this is not the case--- the efficacy of bursty outflows at modifying the DM density depends on factors such as the number of outflow events and, in the case of one burst, the timing of the outflow.

In Figure \ref{varnum}, we see that even with a fixed total outflow mass, both the core radius and the inner log-slope strongly depend on the number of bursts in a non-trivial way.
Initially, when decreasing the bursts from 62 to 5, the dominant effect in terms of the strength of core formation is the amount of mass being expelled: fewer, larger bursts are more impactful.
Critically, we find that this trend reverses as the number of bursts is decreased further. For fewer than five bursts, the dominant factor in determining how many of the inner-region DM particles can be impacted is the number of bursts, not the size of them. 

In addition to showing core formation in the bright dwarf regime,~\citeauthor{Azartash-Namin2024}~\citeyear{Azartash-Namin2024} found that the DM density profiles remain cusped for the UFDs within their sample despite having bursty star formation histories (SFHs) at early times.
This result was independent of the sub-grid feedback model employed.
We similarly find challenges to forming cores in a single (or small number of) early outflows in the UFD regime, as demonstrated in Figure \ref{m8single}.
Our prescriptive approach allows us to identify the two-fold origin of this challenge and study it in further detail: both the outflow mass and timing are critical.
A core of appreciable size only forms if the burst occurs at $z < 6$. 
Notably, this result holds regardless of the total amount of mass expelled, even in the case of expelling $2.9 \times 10^{7} M_{\odot}$, which is $> 10\%$ of the halo mass, and $\sim 10^{4} \times M_*~\rm{at}~z=0$.
Even in this case, a burst at $z=5$ forms a core while a burst at $z=6$ does not. 
In particular, the clear differences that we see in the inner log-slopes and core radii for $z>5$ versus $z\le5$ are most likely the result of the major merger that occurs between $z=6~\rm{and}~z=5$ for this galaxy.
Given that the different feedback models compared in ~\citeauthor{Azartash-Namin2024}~(\citeyear{Azartash-Namin2024}) correspond to different mass outflow rates, our finding that the cusp persists for a wide range of mass expelled when star formation is restricted to early times is consistent with their results.

Even though our simple model is agnostic with respect to star formation model and other details of the sub-grid baryonic physics, these results may still be used to constrain sub-grid modeling within cosmological simulations.
Provided there are more than just a few outflows, we find that stronger, more intermittent feedback is preferred for core formation (Figure \ref{varnum}).
As has been shown, star formation must be tied to dense regions to produce galaxies with realistic properties, and a higher gas density threshold for star formation drives stronger outflows by spatially concentrating supernova energy
~\citep[e.g.,][]{Governato2010, Brook2011, Guedes2011}.
Additionally, linking the star formation rate to the local density of molecular hydrogen can both boost feedback~\citep{Christensen2014} and alter the feedback cycle such that suppression of star formation following a localized burst of star formation is stronger~\citep{AgertzKravstov2015}.
In light of these previous studies, our study may then provide support for models which impose high density thresholds for star formation, either directly or by tying star formation to local molecular hydrogen abundance.

We note that there are some limitations associated with modeling the impact of bursty feedback on the DM particles in this manner:
While we make physical arguments for the timing and magnitude of the bursts in our models, they may not be identical to those that would occur in an observed system. 
We prescribe a set of times for which the central potential is changed dramatically and suddenly, however, nothing physically causes this burst to occur. 
In particular, we do not impose any constraints on the times at which bursts can happen. As described in \S \ref{sec:GrowthModels}, the list of burst times was created by assuming that a burst happens with temporal spacings proportional to the halo dynamical time. One potential consequence of this is that the models with a larger number of bursts 
have bursts occurring later than may be physically realistic. 

The generalizability of our results is limited in two ways. 
Firstly, we do not vary the formation history (in other words, the set of initial conditions) for either halo. 
In the context of the results delineated in \S \ref{subsec:UltraFaintDwarfs}, this may affect the precise range of times for which a single burst can flatten the DM density profile. 
We would not expect this to dramatically affect our qualitative conclusions, though. 
While the transition in core radii/inner log-slope we see in Figure \ref{m8single} is likely a result of the merger that occurs between $z=6~\rm{and}~z=5$, for a galaxy with a different formation history we would expect to find a similar result, where the core does not persist past the most recent major merger.
This is simply because the earlier an outflow occurs, the more likely it is that there is at least one major merger after the outflow.
Secondly, our simulations are of field galaxies, not satellites like the UFDs in the Local Group which, in part, motivated the analysis. 
The role of the environment and the impact that this may have on our results is yet to be tested, though we anticipate that our methods effectively capture the physics of episodic/bursty feedback independent of selected environment.


\section{Conclusions} \label{sec:Conclusions}
In this paper, we introduced a novel approach to modeling the impact of bursty outflows on the orbits of DM particles in the inner region of dwarf galaxies to constrain the regimes for which bursty feedback is capable of turning cusps to cores.
We modeled the gravitational impact of baryons with a massive tracer particle, which allowed us to maintain control over how and when these bursty outflows occur within a cosmological environment.
This technique sits at the intersection of the two avenues bursty feedback has typically been studied through: cosmological simulations where baryonic processes are modeled explicitly (e.g., \cite{Hopkins2014, Onorbe2015, Lazar_2020, Chan_2015}) and the more controlled analytic or idealized modeling of the impact a changing gravitational potential has on the DM particle orbits (e.g., those of \citeauthor{Pontzen_Governato_2012} \citeyear{Pontzen_Governato_2012}). 
Specifically, we introduced this method to evaluate the ability of a single burst to form cores under realistic conditions for an UFD galaxy with a key constraint: the stars form in a short period of time and in the early universe. 
We introduced two suites of modified DMO simulations: a classical dwarf and UFD analog, and studied how varying the prescribed evolution of the tracer particle (representing the baryon mass) in a way that corresponds to different star formation histories impacted the present-day DM density profile.
Our key findings are as follows:
\begin{itemize}
    \item Our simplified model reproduced the cusp-to-core transformation for a galaxy of halo mass $\sim 10^{10} M_{\odot}$, which is the regime where we expect this to take place based on the results of hydrodynamic simulations that model baryonic physics (including bursty feedback) explicitly~\citep{Chan_2015}. We find our smooth-growth model produced a cusp, while many of our bursty models produced cores.
    \item Whether or not a core forms depends on the amount of mass expelled in the burst and how many bursty outflows there are. When the total outflow mass over all bursts is variable, a larger number of bursts or a larger outflow mass produced a more cored profile at present-day. This result was consistent across both metrics that we used to quantify the strength of core formation: the inner log-slope ($\alpha$) and the best-fit value for the core radius ($r_{\rm c}$).
    \item Holding the total mass expelled over all the bursts fixed (and therefore holding the total energy transferred irreversibly to the DM particles fixed), we still found a dependence on the number of bursts that the mass is expelled over. We found that for five or more bursts, it is more effective to have fewer, larger bursts. However, for fewer than five bursts, the profiles became cusped once again even for extremely large outflows. 
    \item Applying this method to an ultra-faint galaxy analogous to the satellites within our Local Group, we found that a single burst was insufficient to transform the DM density profile if we made realistic assumptions about the outflow mass and when the burst occurs. Namely, if we assume a gas fraction of 95\%, that all gas is expelled in the outflow, and that the burst happens prior to $z=6$, we find that the density profile remains cusped.
    \item By varying the timing and outflow mass of our single-burst models, we discovered two barriers to core formation in UFDs that form in one, early burst: (i) the amount of mass that can is ejected must be sufficiently large to impact the DM density, and (ii) the outflow must occur sufficiently late ($z < 6$).
    \item Given that the SFHs of the Local Group satellites are largely constrained to before reionization~\citep{Sacchi2021}, the previous point indicates that a single burst of star formation is insufficient to explain the density profiles of these systems if they are cored as suggested in~\citeauthor{Almeida2024}~\citeyearpar{Almeida2024}. 
\end{itemize}



\section{Acknowledgments}
The authors thank the University of Florida Research Experience for Undergraduates (REU) in Computational and Data-Intensive Astrophysics, funded by the National Science Foundation (NSF-AST 1851954) for supporting this project.
The authors acknowledge UFIT Research Computing for providing computational resources and support that have contributed to the research results reported in this publication.
AMG and PT acknowledge support from NSF-AST 2346977.
PT acknowledges support from NASA ATP grant 80NSSC22K0716. 
This work was supported by the National Science Foundation under Cooperative Agreement 2421782 and the Simons Foundation award MPS-AI-00010515.
ML is supported by the Department of Energy (DOE), under Award Number DE-SC0007968, as well as the Simons Investigator in Physics Award.
\software{Astropy~\citep[][]{apy1, apy2, apy3}},
          Matplotlib \citep{mpl},
          \textsc{colossus} \citep{colossus}, \textsc{arepo} \citep{Springel2010, Weinberger2020}, \textsc{music} \citep{Hahn2011}

\section{Data Availability}
The code used to produce the results of this paper is available on Github\footnote{\url{https://github.com/oliviajmostow/DMOCoreFormation}.} and Zenodo\footnote{Version 2:\dataset[10.5281/zenodo.17227115]{https://doi.org/10.5281/zenodo.17227115}; all versions: \href{https://doi.org/10.5281/zenodo.15856691}{10.5281/zenodo.15856691}.}, and the raw simulation data is accessible via Globus\footnote{\url{https://tinyurl.com/DMOCoreFormation}.}.


\bibliography{example}{}
\bibliographystyle{aasjournal}



\end{document}